\documentclass[10pt, letterpaper]{article}
\usepackage[top=0.85in,left=.8in,footskip=0.75in,marginparwidth=2in]{geometry}

\usepackage{cite}
\usepackage{amsmath}
\usepackage{lineno}
\usepackage{nameref,hyperref}

\usepackage{microtype}
\DisableLigatures[f]{encoding = *, family = * }

\usepackage{changepage}
\usepackage[aboveskip=1pt,labelfont=bf,labelsep=period,singlelinecheck=off]{caption}

\makeatletter
\renewcommand{\@biblabel}[1]{\quad#1.}
\makeatother

\usepackage{soul}
\usepackage{graphicx}
\usepackage{hyperref}
\usepackage{placeins}
\usepackage{mathptmx}
\usepackage{anyfontsize}
\usepackage{t1enc}
\usepackage{lastpage,fancyhdr,graphicx}
\usepackage{epstopdf}
\usepackage{cite} 
\usepackage{multirow}
\usepackage{url}
\usepackage{tabularx}
\usepackage{placeins}
\usepackage{multicol,lipsum}
\usepackage{rotating}
\usepackage{natbib}
\usepackage{float}
\usepackage{booktabs}

\hypersetup{colorlinks,allcolors=black}
\fancyheadoffset[L]{2.25in}
\fancyfootoffset[L]{2.25in}

\usepackage{color}
\usepackage{float}

\definecolor{Gray}{gray}{.25}

\usepackage{graphicx}

\usepackage{sidecap}

\usepackage{wrapfig}
\usepackage[pscoord]{eso-pic}
\usepackage[fulladjust]{marginnote}
\reversemarginpar

\begin{document}
	\vspace*{0.35in}
	
	\begin{flushleft}
		{\Large
			\textbf\newline{Innovations in Surface Modification Techniques: Advancing Hydrophilic \textit{LiYF$_{4}$:Yb, Er, Tm}  Upconversion Nanoparticles and Their Applications}
		}
		\newline
		\\
            Shahriar Esmaeili\textsuperscript{1, 2},
		Navid Rajil\textsuperscript{1, 2},
		Masfer H. Alkahtani\textsuperscript{3},
		Yahya A. Alzahrani\textsuperscript{3},
            Ayla Hazrathosseini \textsuperscript{1, 2},
		Benjamin W. Neuman\textsuperscript{4},
            Zhenhuan Yi\textsuperscript{1, 2},
            Robert W. Brick \textsuperscript{1, 2},
            Alexei V. Sokolov\textsuperscript{1, 2, 4},
            Philip R. Hemmer \textsuperscript{1, 2, 4}, and
            Marlan O. Scully \textsuperscript{1, 2}

		\bigskip
	
	\textsuperscript{1}Institute for Quantum Science and Engineering, Texas A\&M University, College Station, TX 77843, USA\\
 \textsuperscript{2}Dept. of Physics and Astronomy, Texas A\&M University, College Station, TX 77843, USA\\
\textsuperscript{3}King Abdulaziz City for Science and Technology (KACST), Riyadh 11442, Saudi Arabia\\
\textsuperscript{4}Dept. of Biology, Texas A\&M University, College Station, TX 77843, USA\\
\textsuperscript{5}Dept. of Electrical and Computer Engineering, Texas A\&M University, College Station, TX 77843, USA\\

		\bigskip

	\end{flushleft}
	\justify
	\section*{Abstract}

The development and application of upconversion nanoparticles (UCNPs) have garnered significant attention due to their unique optical properties and potential uses in bioimaging, drug delivery, and solar cells. However, the hydrophobic nature of UCNPs presents challenges in their synthesis and application, particularly in aqueous environments. We provide an overview of UCNPs, their synthesis challenges, and the importance of surface modification. Furthermore, we discuss the properties of \textit{LiYF$_{4}$:Yb, Er, Tm}  UCNPs synthesized using novel 2,2’-[ethylenebis(oxy)] bisacetic acid (EBAA) method and their versatile applications. Notably, the first Dynamic Light Scattering measurement on 05/22/2022 showed a size of 11.39 nm, and after 348 days on 04/05/2023, the same batch maintained a size of 13.8 nm, indicating excellent stability and no particle agglomeration over this extended period. This remarkable stability underscores the potential of UCNPs synthesized with the EBAA method for long-term applications. Finally, we compare the EBAA method with other surface modification techniques, exploring challenges and future perspectives for the use of hydrophilic UCNPs in various applications. This review aims to emphasize the significance of the EBAA method in advancing the field of upconversion nanoparticles and broadening their potential integration into diverse applications.
\\
\newline
	\textbf{Keywords:} 
	upconversion nanoparticles, hydrophilic UCNPs, 2,2'-[ethylenebis(oxy)] bisacetic acid (EBAA), surface modification, biosensing, quantum sensing, solar cells, stability, biocompatibility
	
	\section{Introduction}\label{sec1}
Upconversion nanoparticles (UCNPs) are luminescent nanomaterials capable of converting lower-energy photons into higher-energy photons through a nonlinear optical process known as photon upconversion \citep{zhou2015upconversion, haase2011upconverting}. This process typically involves the absorption of two or more lower-energy photons, followed by the emission of a single higher-energy photon \citep{chen2014upconversion}. Exhibiting sharp emission bands, low autofluorescence background, high photostability, and long luminescence lifetimes, UCNPs are ideal for a wide array of applications \citep{wang2009recent}.

In bioimaging, UCNPs have demonstrated potential in fluorescence microscopy, single-molecule imaging, and deep-tissue imaging, attributable to their high signal-to-noise ratio, low toxicity, and deep tissue penetration \citep{cheng2013upconversion, li2018upconversion}. For drug delivery, they serve as carriers, leveraging unique optical properties for photo-triggered release and real-time monitoring of drug release kinetics \citep{wang2013upconversion, liu2015silica}. In solar cells, they hold the promise of enhancing photovoltaic device efficiency by converting low-energy photons into higher-energy ones absorbable by solar cell materials, thus boosting overall conversion efficiency \citep{goldschmidt2015upconversion, zhang2021lanthanide}.

The hydrophobicity of upconversion nanoparticles (UCNPs) is influenced by the nature of their host materials and the prevalent capping agents used during their synthesis. Yttrium lithium fluoride (YLF), a laser crystal commonly used for UCNPs, is known for its superior optical properties and stable crystal surface structures which inherently repel water due to their chemistry. Additionally, the prevalent use of oleate ligands during the synthesis process further contributes to the hydrophobic nature of UCNPs, as these ligands form a robust organic layer on the nanoparticle surface through hydrophobic interactions, shielding the core from aqueous environments. While this oleate capping is crucial for preserving the structural and optical integrity of UCNPs, it presents a significant barrier to their direct use in water-based applications, necessitating innovative surface modification approaches to replace the hydrophobic oleate layer with hydrophilic groups \citep{kong2017general, zhou2015upconversion}.

However, the synthesis and surface modification of UCNPs present challenges that must be overcome to fully harness their potential. A primary challenge involves rendering UCNPs hydrophilic since they are inherently hydrophobic and prone to aggregation in aqueous solutions, leading to fluorescence quenching and diminished performance \citep{xie2013mechanistic}. Surface modification is essential to improve stability, biocompatibility, and optical properties, crucial for successful integration into various applications, particularly in biomedicine \citep{gnach2012lanthanide}. Reported surface modification techniques for UCNPs, such as ligand exchange, covalent functionalization, and encapsulation, often face limitations in stability, reproducibility, and toxicity \citep{chen2018selective}. Furthermore, certain modification strategies can produce a thick coating, potentially hindering the interaction with target molecules or particles, thus affecting performance in some applications \citep{arppe2015quenching}.

A comprehensive review by \cite{jiang2022comprehensive} highlights the diverse applications of near-infrared-excited upconversion nanoparticles (UCNPs) across environmental, bioscience, food science, and medical domains. This review underscores UCNPs' multicolor emissions, low auto-fluorescence, high chemical stability, and extended fluorescence lifetime. It also emphasizes the need for tailored luminescence efficiency and the design of versatile UCNP analysis platforms \citep{jiang2022comprehensive}. In alignment with these directions, a targeted study by Zhang et al. \citep{zhang2009hexanedioic} addresses UCNPs' hydrophilicity enhancement for biological applications. Their work demonstrates the successful conversion of hydrophobic NaYF$_4$:Yb, Er nanoparticles into hydrophilic ones using Hexanedioic acid (HDA) through a ligand exchange process \citep{zhang2009hexanedioic}.

The synergy between \cite{jiang2022comprehensive}'s comprehensive review and \citep{zhang2009hexanedioic}'s targeted study exemplifies the collaborative effort needed to address existing challenges, setting the stage for broader applications and enhanced effectiveness of UCNPs in various fields. In response to the challenges outlined, a novel synthesis method for hydrophilic UCNPs using 2,2'-[ethylenebis(oxy)] bisacetic acid (EBAA) as a surface modifier has been proposed. This innovative approach circumvents some of the limitations of traditional surface modification techniques by offering a thin yet effective coating that fosters interaction with other particles or molecules while also preventing water-induced quenching of particle fluorescence \citep{wu2015upconversion}. The EBAA method has been promising in yielding hydrophilic UCNPs with enhanced stability, biocompatibility, and optical properties, rendering them more amenable for diverse applications, especially in biosensors.

This review endeavors to present a comprehensive overview of the EBAA method for hydrophilic UCNPs synthesis, delineating its advantages over existing surface modification techniques and potential impacts across various applications. The discussion commences with an array of hydrophilic UCNPs applications, concentrating on their roles in biosensing, bioimaging, drug delivery, and solar cells. It then delves into UCNPs' synthesis background, addressing surface modification challenges. Subsequently, the EBAA method's key features are introduced, followed by an in-depth analysis of the properties of hydrophilic UCNPs synthesized via this method. Additionally, a comparative perspective of the EBAA method against other surface modification techniques is provided, coupled with an exploration of challenges and future perspectives for hydrophilic UCNPs applications. The review culminates by accentuating the EBAA method's significance in propelling the field of upconversion nanoparticles forward and unlocking new avenues for their integration into a multitude of applications.

\section{Applications of Hydrophilic Upconversion Nanoparticles}\label{sec3}

Now, we explore the various applications of hydrophilic UCNPs, focusing on their use in biosensing, bioimaging, drug delivery, and solar cells.

\subsection{Biosensing}\label{sec2.1}

Hydrophilic UCNPs exhibit good dispersibility in aqueous environments, which is critical for their successful integration into biological systems.  Various biosensing applications of hydrophilic UCNPs, including the detection of nucleic acids, proteins, and small molecules, as well as their use in cellular and in vivo imaging are discussed as follows;

\begin{itemize}
    \item \textbf{Nucleic Acid Detection:}
Hydrophilic UCNPs have been employed in the detection of nucleic acids, such as DNA and RNA, through fluorescence resonance energy transfer (FRET) mechanisms \citep{bhuckory2023understanding}. In these systems, UCNPs serve as energy donors, while fluorophores or other chromophores act as energy acceptors. By functionalizing the surface of hydrophilic UCNPs with specific oligonucleotide probes, it is possible to achieve highly sensitive and selective detection of target nucleic acid sequences, such as viral or bacterial DNA/RNA, (Fig. \ref{fig6}). In this regard, \cite{esmaeili2022detection, esmaeili2023advances} has also proposed to utilize hydrophilic UCNPs synthesized via the EBAA method for the detection of SARS-CoV-2 cDNA using FRET. This approach demonstrates the potential of hydrophilic UCNPs in the development of rapid and accurate diagnostic tools for various diseases.

 \begin{figure}[H]
		\centerline{\includegraphics[width=1.1\textwidth,clip=]{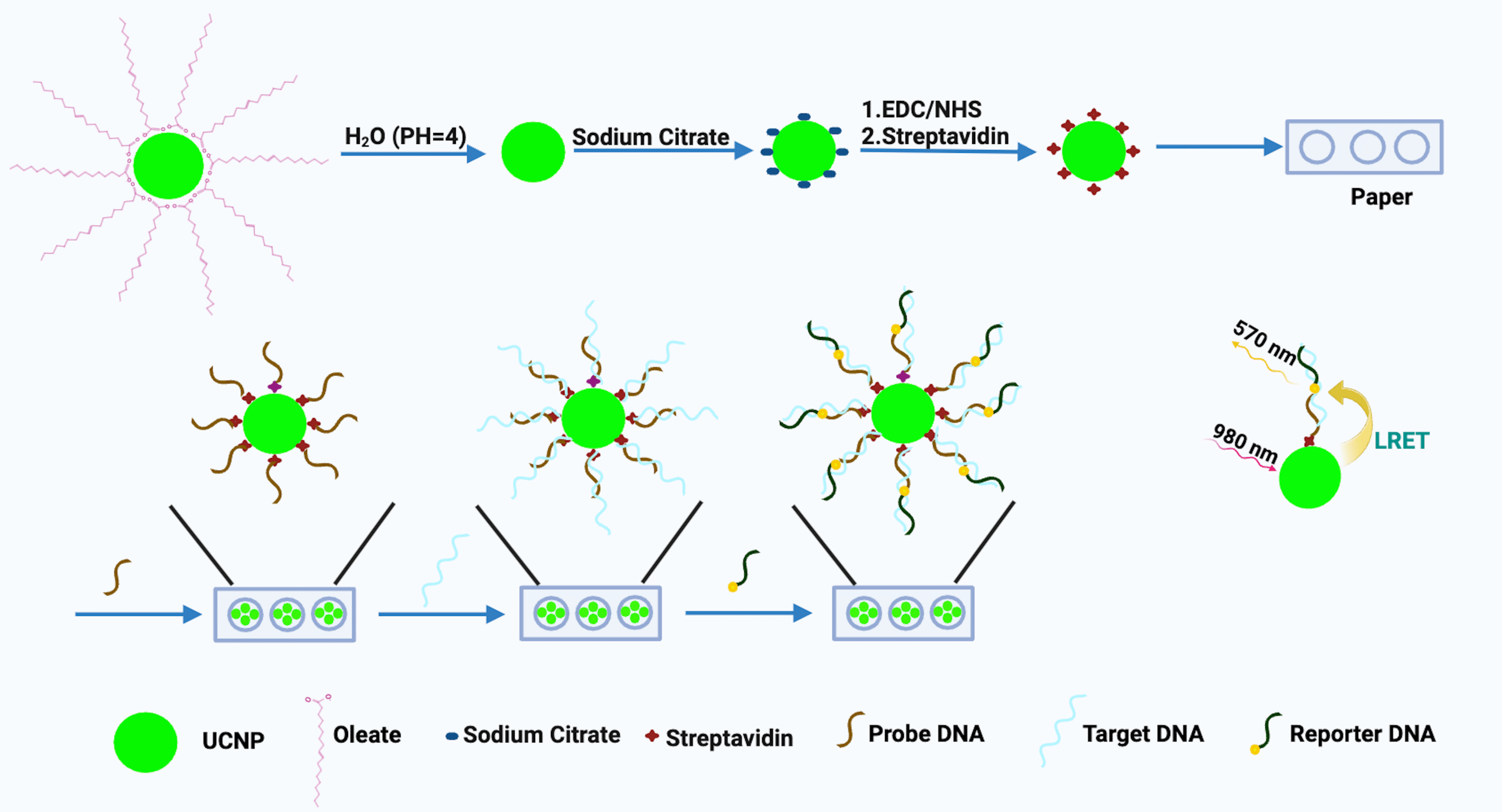}}
		\caption{A schematic diagram showing a sandwich format DNA assay on paper that uses hydrophilic upconversion nanoparticles (UCNPs) as energy donors in luminescence resonance energy transfer (LRET).}\label{fig6}
	\end{figure}

\item \textbf{Protein Detection:}
The detection of proteins is another important application of hydrophilic UCNPs in biosensing. By conjugating UCNPs with specific antibodies or other recognition elements, it is possible to design highly selective and sensitive protein biosensors \citep{cao2022assessing}. These sensors can be used to detect and quantify the presence of specific proteins in complex biological samples, such as blood, serum, or cellular extracts. For instance, hydrophilic UCNPs have been utilized to detect cancer biomarkers, enabling the early diagnosis of various types of cancers \citep{chinen2015nanoparticle}. Additionally, bioconjugated hydrophilic UCNPs have been employed in the detection of enzymes, offering a versatile platform for monitoring enzymatic activity and inhibition \citep{rajil2022quantum, rajilsupplementary}.

 \begin{figure}[H]
		\centerline{\includegraphics[width=1\textwidth,clip=]{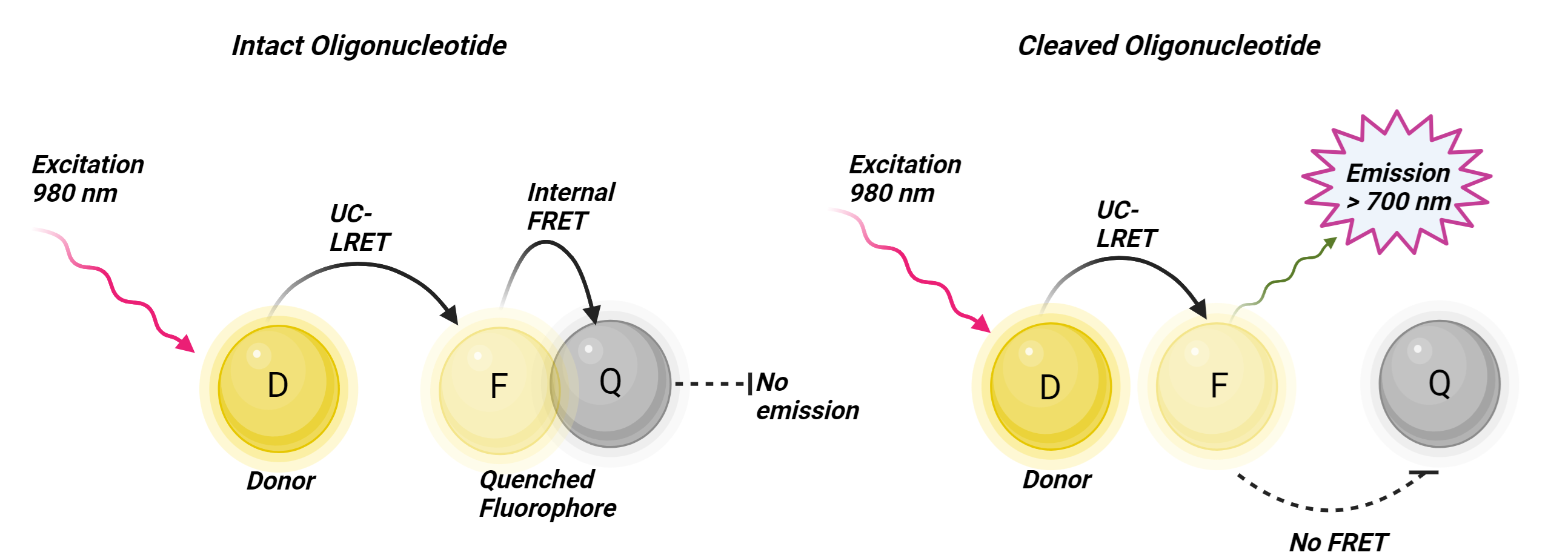}}
		\caption{This figure illustrates the basic concept of a homogeneous enzyme activity assay system, where a hydrolytic enzyme reaction leads to the separation of a fluorophore (F) and a quencher (Q) located at opposite ends of an oligonucleotide, resulting in the emission of the fluorophore (measured at >700 nm).}\label{fig7}
	\end{figure}

 \item \textbf{Small Molecule Detection:}

 Hydrophilic UCNPs have also been used to develop biosensors for the detection of metal ions, anions, small molecules and other biologically relevant compounds \citep{su2021paa}. By functionalizing the surface of UCNPs with selective receptors or ligands, it is possible to achieve high selectivity and sensitivity in detecting specific small molecules. Applications of these biosensors range from environmental monitoring to clinical diagnostics and drug discovery. For example, hydrophilic UCNPs have been utilized to detect heavy metal ions in water samples, offering a rapid and cost-effective method for monitoring water quality \citep{guo2016sensitive}, see Fig. \ref{fig8}.

  \begin{figure}[H]
		\centerline{\includegraphics[width=0.5\textwidth,clip=]{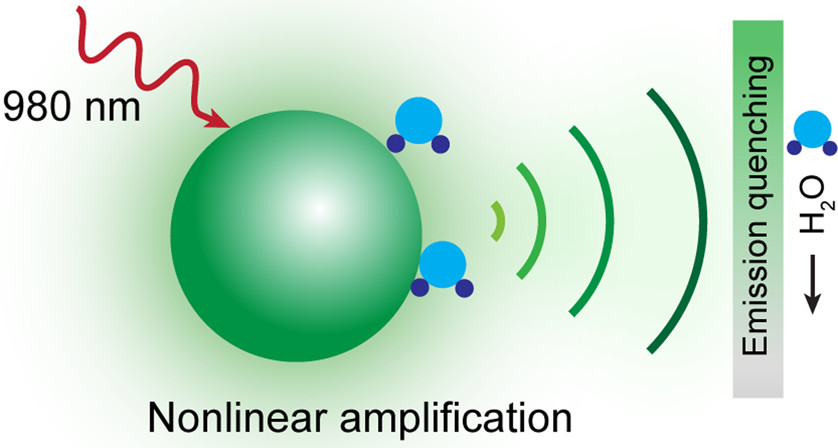}}
		\caption{Schematic of upconversion emission quenching by water molecules, which absorb strongly at 980 nm. The figure implies the energy absorption by water leading to emission quenching. Adapted from \citep{guo2016sensitive}.}\label{fig8}
	\end{figure}

\item \textbf{Cellular and In Vivo Imaging}

Finally, hydrophilic UCNPs have been applied in cellular and in vivo imaging, where they serve as luminescent probes for the visualization of biological structures and processes. Owing to their unique upconversion properties and high photostability, UCNPs can provide high-resolution and real-time imaging of living cells and organisms, with minimal phototoxicity and autofluorescence background. This enables the study of various biological phenomena, such as gene expression, protein-protein interactions, and cellular signaling pathways, with unprecedented detail and precision (Fig. \ref{fig9}).

  \begin{figure}[H]
		\centerline{\includegraphics[width=0.75\textwidth,clip=]{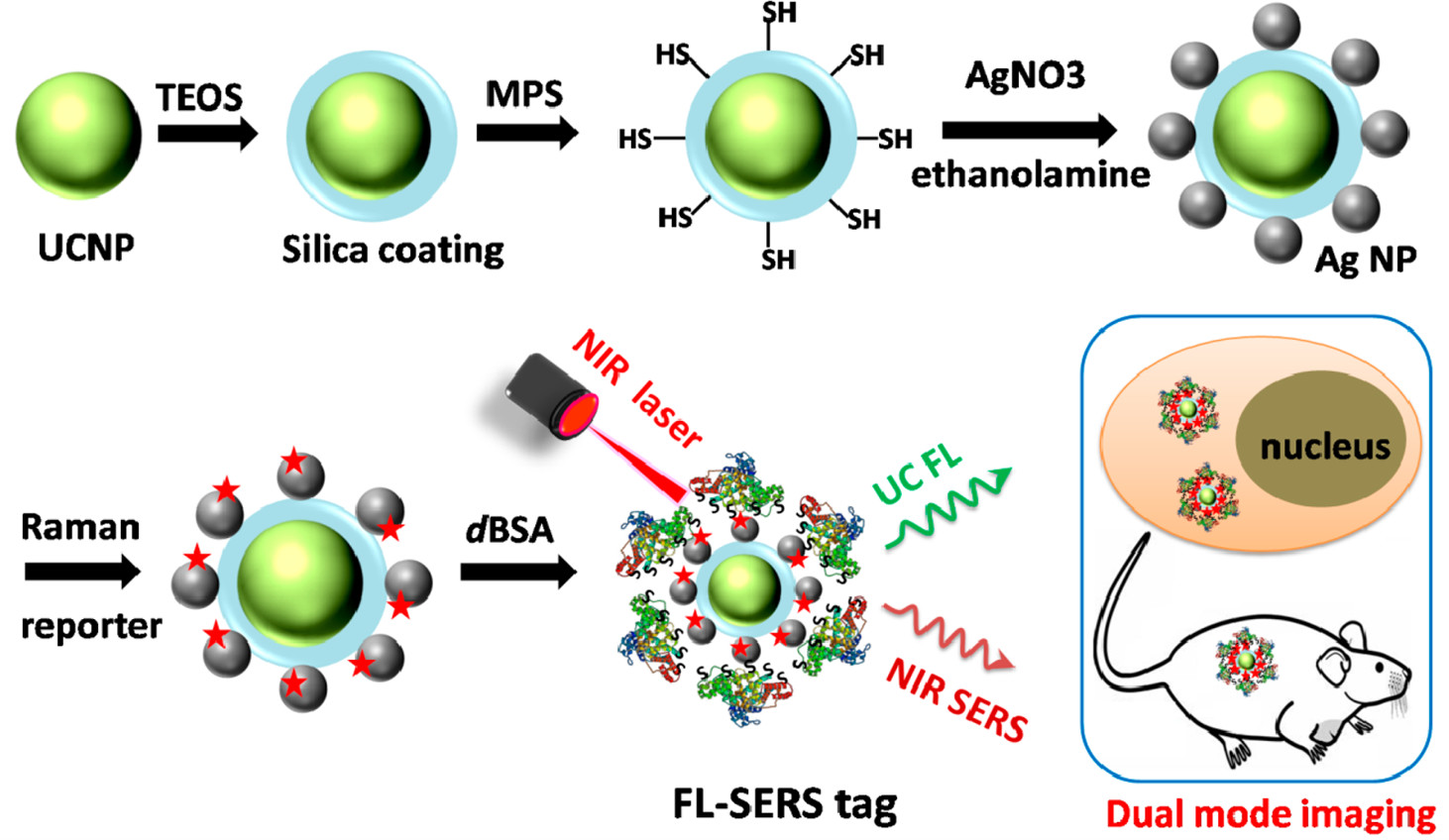}}
		\caption{Schematic illustration of UCF-SERS Dots and the demonstration of the \textit{in vivo} imaging capabilities of UCF-SERS  on living mice. Adapted from \citep{niu2014upconversion}.}\label{fig9}
	\end{figure}

\end{itemize}

\subsection{Bioimaging}\label{sec2.2}

\begin{figure}[H]
		\centerline{\includegraphics[width=0.75\textwidth,clip=]{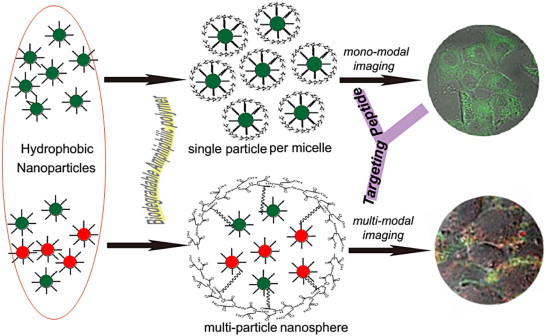}}
		\caption{Illustration of targeted bioimaging using hydrophilic UCNPs functionalized with biomolecules and their excitation by NIR light. Adapted from \citep{xu2015upconversion}.}\label{fig3}
	\end{figure}

Luminescent bioimaging with hydrophilic UCNPs enables high-resolution, real-time imaging of biological structures like cells and tissues with minimal invasiveness \citep{zhou2015upconversion}. These nanoparticles can be selectively targeted to specific cellular markers by functionalizing them with biomolecules, such as antibodies or peptides \citep{wang2009recent, rajil2021fiber, hemmer2023opportunities}. Targeted UCNPs, when excited by near-infrared (NIR) light, penetrate tissues deeply with less photodamage than ultraviolet or visible light, resulting in efficient upconversion emission for high-contrast images. Although IR light typically encounters water absorption issues, the NIR wavelengths used for UCNPs fall within the biological tissue 'optical window,' where water's absorption is significantly lower, thus mitigating this problem and allowing effective tissue penetration. This has been demonstrated in various applications, such as in vivo tumor imaging \citep{zhan2011using}, stem cell tracking \citep{xiong2009high}, and neural activity visualization \citep{liu2021near}. UCNPs enhance optical coherence tomography (OCT) and non-invasive imaging due to their photophysical properties and ability to convert NIR excitation to visible or ultraviolet emissions. Incorporating UCNPs as contrast agents or biosensing probes can thus improve the sensitivity and specificity of OCT imaging and biosensing in biomedical fields including ophthalmology, oncology, and cardiovascular research (Fig. \ref{fig3})\citep{sen2021extended, sen2022implementation, classen2022modeling}.

\subsection{Drug delivery}\label{sec2.3}

\begin{figure}[H]
		\centerline{\includegraphics[width=0.65\textwidth,clip=]{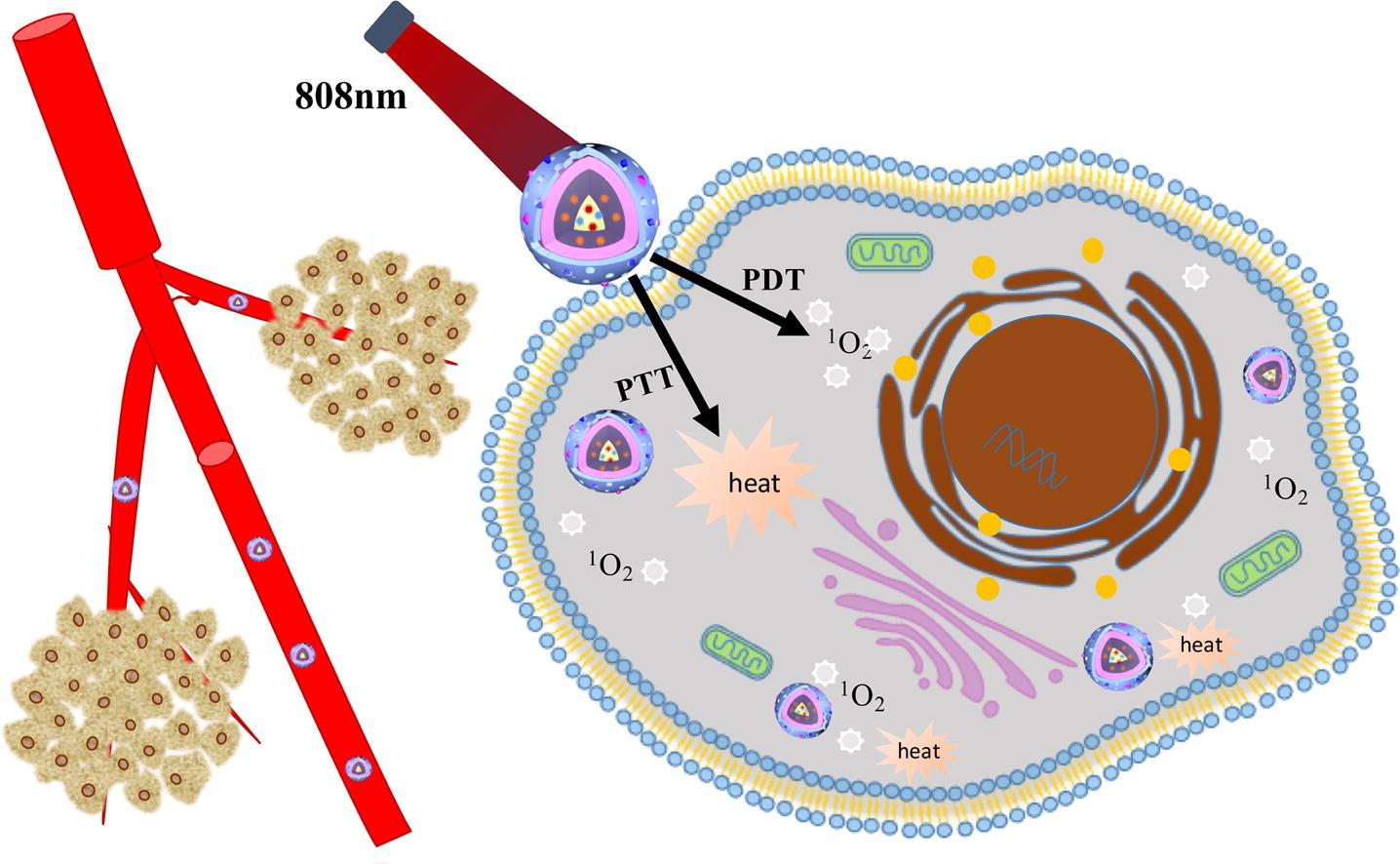}}
		\caption{Schematic illustration of the application of upconversion nanoparticles (UCNPs) in photodynamic therapy (PDT) and photothermal therapy (PTT). Hydrophilic UCNPs are targeted to cancer cells. Upon near-infrared (NIR) excitation, UCNPs convert the low-energy NIR light to higher-energy visible or UV light, activating the photosensitizers in PDT or producing localized heat in PTT, leading to cancer cell destruction. Adapted from \citep{zhang2021synergistic}.}\label{fig4}
	\end{figure}

Hydrophilic UCNPs have also shown great potential in drug delivery systems. Their unique optical properties enable controlled and targeted drug release, while their hydrophilic nature ensures biocompatibility and effective dispersion in biological systems. UCNPs can be used as drug carriers by encapsulating or conjugating therapeutic agents to their surfaces \citep{muhr2014upconversion}. The upconversion luminescence of these nanoparticles can be utilized to trigger drug release through photodynamic therapy (PDT) or photothermal therapy (PTT). In PDT, the UCNPs can generate reactive oxygen species upon NIR light excitation, leading to the destruction of cancer cells \citep{lucky2015nanoparticles, idris2012vivo}. In PTT, the UCNPs can convert absorbed NIR light into heat, causing local hyperthermia and subsequent drug release \citep{lv2017situ}, see Fig. \ref{fig4}.

\subsection{Solar cells}\label{sec2.4}

The application of upconversion nanoparticles (UCNPs) in solar cells has emerged as a promising strategy to enhance the performance of photovoltaic devices. By converting low-energy photons from the solar spectrum into high-energy photons, UCNPs can increase the utilization of solar energy and, consequently, the overall efficiency of solar cells. Traditional solar cells, such as silicon-based devices, have limited spectral response due to their specific bandgap, which restricts the absorption of photons with energy lower than the bandgap. As a result, a large portion of the solar spectrum remains unutilized, leading to suboptimal energy conversion efficiency. To overcome this limitation, researchers have explored the use of UCNPs to increase the utilization of the solar spectrum \citep{trupke2002improving}.

UCNPs can absorb multiple low-energy photons (usually in the near-infrared (NIR) range) and convert them into a single high-energy photon (typically in the visible or ultraviolet range) through a process called upconversion. By incorporating UCNPs into solar cells, researchers aim to capture and convert the low-energy photons that would otherwise be wasted. The upconverted high-energy photons can then be absorbed by the solar cell, generating additional photocurrent and improving the overall energy conversion efficiency \citep{huang2013enhancing}.

Several strategies have been employed to integrate UCNPs into solar cells, such as embedding them into the active layer or the rear reflector, incorporating them into luminescent down-shifting layers, or employing them in luminescent solar concentrators. These approaches are applicable to a wide array of solar cells, including silicon-based devices, dye-sensitized solar cells, organic photovoltaics, and perovskite solar cells \citep{atabaev2019upconversion}. The integration of hydrophilic UCNPs into solar cells offers distinct advantages. Their compatibility with aqueous environments and water-based processing techniques can streamline the incorporation into various architectures, potentially reducing fabrication costs and environmental impacts. Moreover, their enhanced colloidal stability and reduced aggregation can lead to more efficient upconversion, thereby improving solar cell performance.

\begin{figure}[H]
		\centerline{\includegraphics[width=0.75\textwidth,clip=]{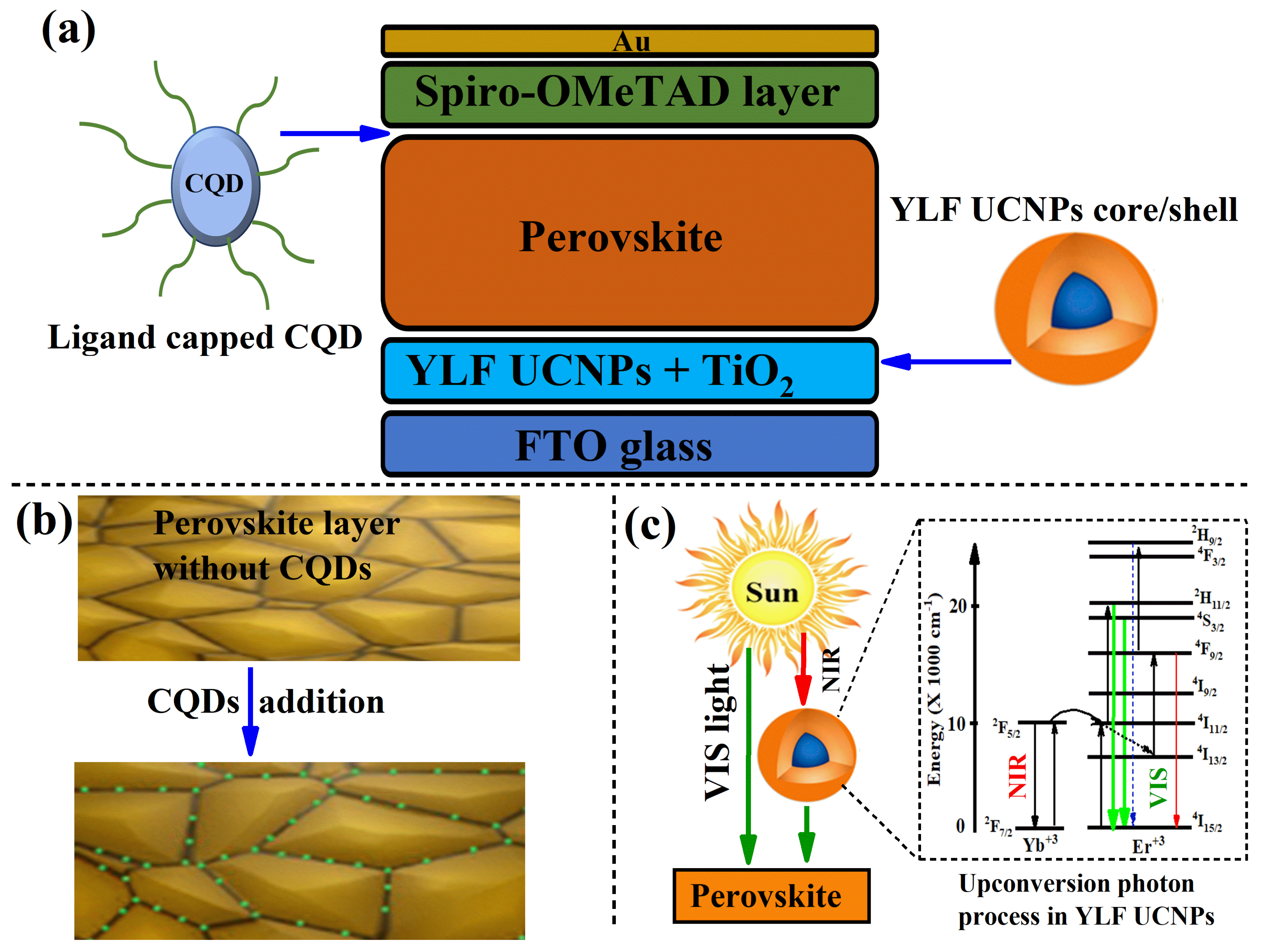}}
		\caption{(a) Showig the use of ligand-capped carbon quantum dots (CQDs) and UCNPs in perovskite solar cells (PSCs). The ligand-capped CQDs are mixed at different concentrations with the perovskite layer, while the UCNPs are embedded in the mesoporous layer. (b) The proposed passivation mechanism of CQDs in the PSC devices is also illustrated. (c) Additionally, the figure depicts the upconversion process of converting wasted NIR photons of sunlight to visible light by the UCNPs. Adapted from \citep{alkahtani2022high}.}\label{fig5}
	\end{figure}

The capability of hydrophilic UCNPs to convert low-energy, non-absorbed photons into higher-energy photons absorbable by solar cell materials is a pivotal advantage. This upconversion process can enhance the light-harvesting capabilities of solar cells, especially in the NIR region, where conventional cells exhibit limited absorption. Research exploring the integration of UCNPs, particularly hydrophilic variants, into diverse solar cells like dye-sensitized solar cells (DSSCs) \citep{liu2021enhanced}, quantum dot-sensitized solar cells (QDSSCs) \citep{sun2018upconverting}, and perovskite solar cells \citep{alkahtani2022high}, has shown promising results. As illustrated in Fig. \ref{fig5}, incorporating hydrophilic UCNPs can significantly enhance power conversion efficiencies. Their stability and reduced tendency to aggregate further contribute to the improved performance of the solar cells \citep{hao2017enhancing}.

In summary, UCNPs, particularly hydrophilic UCNPs, hold great potential for enhancing the performance of solar cells by increasing the utilization of the solar spectrum through upconversion. The integration of UCNPs into solar cells can pave the way for the development of more efficient and sustainable photovoltaic technologies \citep{shang2015enhancing, yu2014enhanced, li2011core}.

\section{UCNPs: Synthesis Challenges and the Importance of Surface Modification} \label{sec2}

\begin{figure*}[ht]
		\centerline{\includegraphics[width=.75\textwidth,clip=]{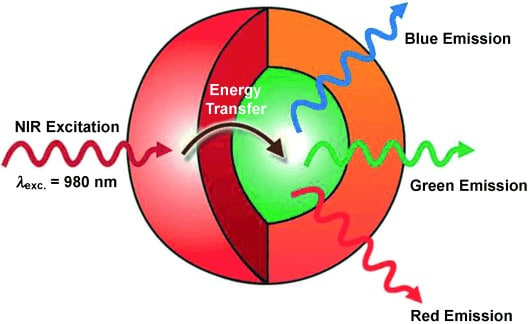}}
		\caption{A schematic representation of the upconversion process in lanthanide-doped UCNPs, illustrating the sequential absorption of multiple low-energy photons and the emission of a single high-energy photon. Adapted from \citep{vetrone2009active}.}\label{fig1}
	\end{figure*}

UCNPs are distinguished by their ability to convert low-energy photons into higher-energy emission, an upconversion process not seen in traditional luminescent materials \citep{chen2014upconversion}, as illustrated in Fig. \ref{fig1}. Their sharp emission bands, significant anti-Stokes shifts, and resistance to photobleaching underscore their potential for advancing bioimaging, drug delivery, and solar cell technologies. 

The synthesis of UCNPs typically involves the incorporation of lanthanide ions, such as ${Yb}^3+,~{Er}^3+, ~and~{Tm}^3+$, into a host matrix, usually an inorganic material like NaYF$_4$, NaGdF$_4$, and LiYF$_4$ through a variety of methods, including hydrothermal synthesis, solvothermal synthesis, and co-precipitation. The choice of lanthanide ions and the composition of the host matrix greatly influence the upconversion efficiency and the emission spectrum of the resulting nanoparticles \citep{wang2010upconversion}. However, the synthesis of UCNPs presents several challenges, such as the formation of non-luminescent surface defects, size and shape control, and the need for high-temperature annealing to achieve efficient upconversion \citep{liu2011sub}. Moreover, the surface of as-synthesized UCNPs is often hydrophobic, which can lead to aggregation and fluorescence quenching when dispersed in water-based solutions, limiting their applicability in aqueous environments \citep{wang2009recent}.

To overcome these challenges, surface modification of UCNPs is essential for enhancing their stability, biocompatibility, and optical properties. Various surface modification techniques have been developed, including ligand exchange, covalent functionalization, and encapsulation \citep{wang2009immunolabeling}. However, these existing techniques often suffer from limitations such as poor stability, low reproducibility, and potential toxicity. Therefore, the development of new methods for synthesizing hydrophilic UCNPs with improved properties is highly desired.


\section{The EBAA Method and Properties of Hydrophilic UCNPs} \label{sec4}

\begin{figure*}[h]
		\centerline{\includegraphics[width=1.2\textwidth,clip=]{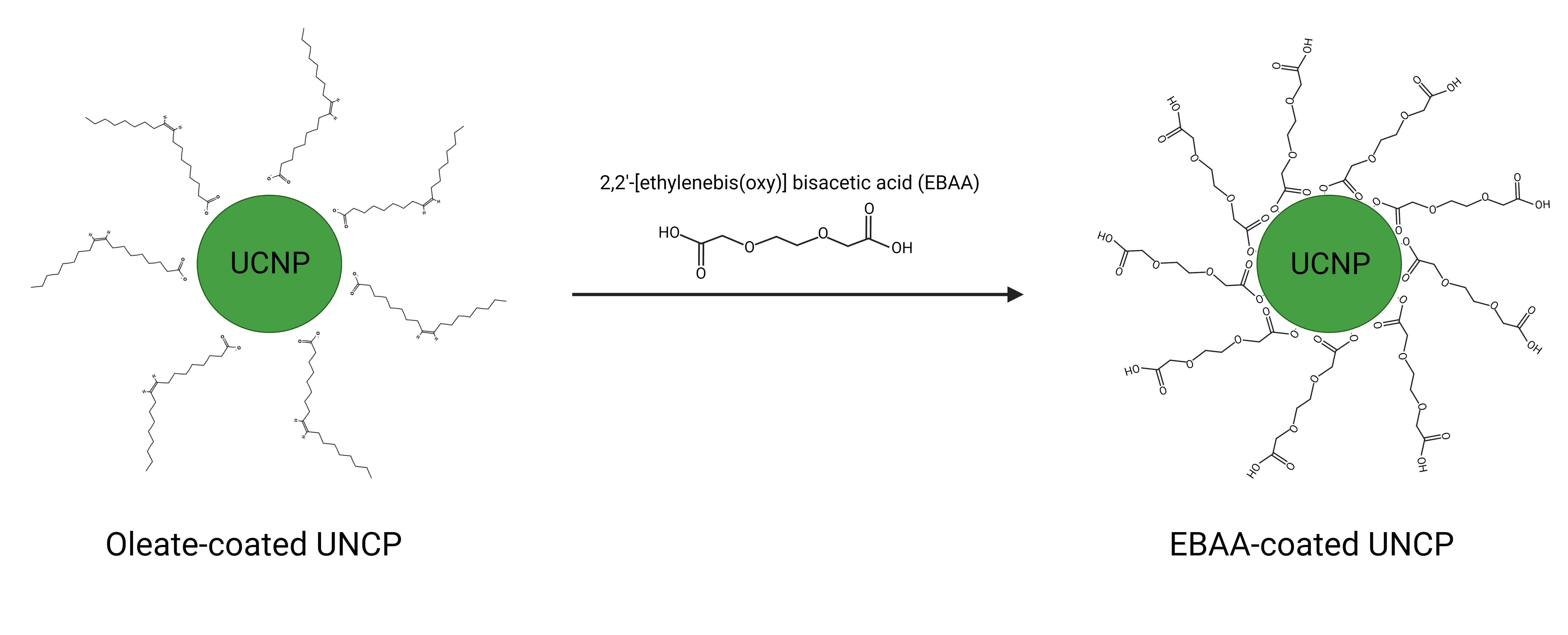}}
		\caption{A schematic representation of the EBAA method, illustrating the ligand exchange process with EBAA, and resulting hydrophilic UCNP.}\label{fig2}
	\end{figure*}

The EBAA (2,2'-[ethylenebis(oxy)] bisacetic acid) method is a novel approach for synthesizing hydrophilic upconversion nanoparticles (UCNPs) that overcomes the challenges associated with traditional surface modification techniques. By using EBAA as a surface modifier, this method enables the synthesis of UCNPs with enhanced stability, biocompatibility, and optical properties, making them more suitable for various applications in aqueous environments \citep{dong2015lanthanide, wang2010upconversion}. The key features of the EBAA method involve the ligand exchange process and dialysis steps. The synthesized hydrophobic UCNPs are dispersed in a solution containing EBAA, which acts as a surface-modifying agent (Fig. \ref{fig2}). 

\subsection{Materials and equipment}\label{sec4.1}
\subsubsection{Reagents}

Oleic acid (technical grade, 90\%; Sigma-Aldrich, cat. no. 364525), 1-Octadecene (technical grade, 90\%; Sigma-Aldrich, cat. no. O806), Methanol (reagent grade), Absolute Ethanol 200 proof (reagent grade), Cyclohexane (reagent grade), Deionized (DI) water, Yb(CH$_3$CO$_2$)$_3$·4H$_2$O (99.9\% trace metals basis; Sigma-Aldrich, cat. no. 326011), Tm(CH$_3$CO$_2$)$_3$·xH$_2$O  (99.9\% trace metals basis; Sigma-Aldrich, cat. no. 367702), Y(CH$_3$CO$_2$)$_3$·xH$_2$O  (99.9\% trace metals basis; Sigma-Aldrich, cat. no. 326046),  Er(CH$_3$CO$_2$)$_3$·4H$_2$O (99.9\% trace metals basis; Sigma-Aldrich, cat. no. 207234-04-6), Lithium acetate (99.95\% trace metals basis, cat. no. 517992), NH$_4$F (ACS reagent, >98\%; Sigma-Aldrich, cat. no. 216011), Dialysis Tubing, 3.5K MWCO, 35 mm (cat. no.: 88244), Diethylene glycol (DEG 99\%, Sigma), 2,2'-[ethylenebis(oxy)] bisacetic acid (cat. no. 23243-68-7). The chemical materials were used as is without any additional purification steps.

\subsubsection{Equipment}
Schlenk line, magnetic stirrer, heating mantle with temperature controller, Argon (99.9\%),
three-neck 100-ml flask, centrifuge, transmission electron microscope (TEM, JEOL 1200), and carbon-coated copper TEM grids.

\subsection{Synthesis of LiYF$_4$:Yb(18\%), Er(1.5\%), Tm(0.5\%) core particles}

At room temperature, 25 mL of oleic acid, and 25 mL of 1-octadecene are added into a 250 mL three-neck flask. The solvent under stirring was degassed using a Schlenk-line at 100 °C and vacuum conditions (~0.8 torr) until the release of gases (air and water) ceased. At room temperature, under a flow of argon gas, a mixture of lithium acetate (10 mmol) and rare earth acetates (5 mmol), including  4 mmol of Y(CH$_3$CO$_2$)$_3$·xH$_2$O, 0.9 mmol of Yb(CH$_3$CO$_2$)$_3$·4H$_2$O, 0.075 mmol of Er(CH$_3$CO$_2$)$_3$·4H$_2$O stock solution,  0.025 mmol of Tm(CH$_3$CO$_2$)$_3$·H$_2$O, are added to the degassed solvent.  

The mixture that was being stirred was once again heated to 100 °C under a vacuum and left at this temperature until the release of gas (acetic acid) stopped. The clear solution was heated to 300°C with argon, and the heating was stopped after 10 minutes. The mixture was then left to cool to 100°C. The solution was degassed by applying a vacuum at 100 °C for one hour to eliminate the acetic acid that was produced during the heating process at 300 °C. Next, the reactor was purged with argon gas, and then 25 mmol of dry NH4F was added to the mixture at 100 °C. The reactor was purged three times with a short vacuum (2-3 mins) and refilled with argon to remove any air. The mixture was stirred at 300°C for 2 hours under an argon atmosphere. After cooling to room temperature, the mixture was centrifuged to obtain a yellowish solution. The resulting particles were collected and subjected to three washes with a 1:1 (v/v) mixture of cyclohexane and ethanol (4 mL). After the final wash, the particles (20 mg/ml) were re-suspended in a new 10 mL solution of cyclohexane and stored at approximately 4 °C \citep{alkahtani2021engineering, carl2021liyf}.

\subsection{Surface–ligand exchange of UCNPs by EBAA (2,2'-[ethylenebis(oxy)] bisacetic acid)}

\begin{figure}[H]
		\centerline{\includegraphics[width=.85\textwidth,clip=]{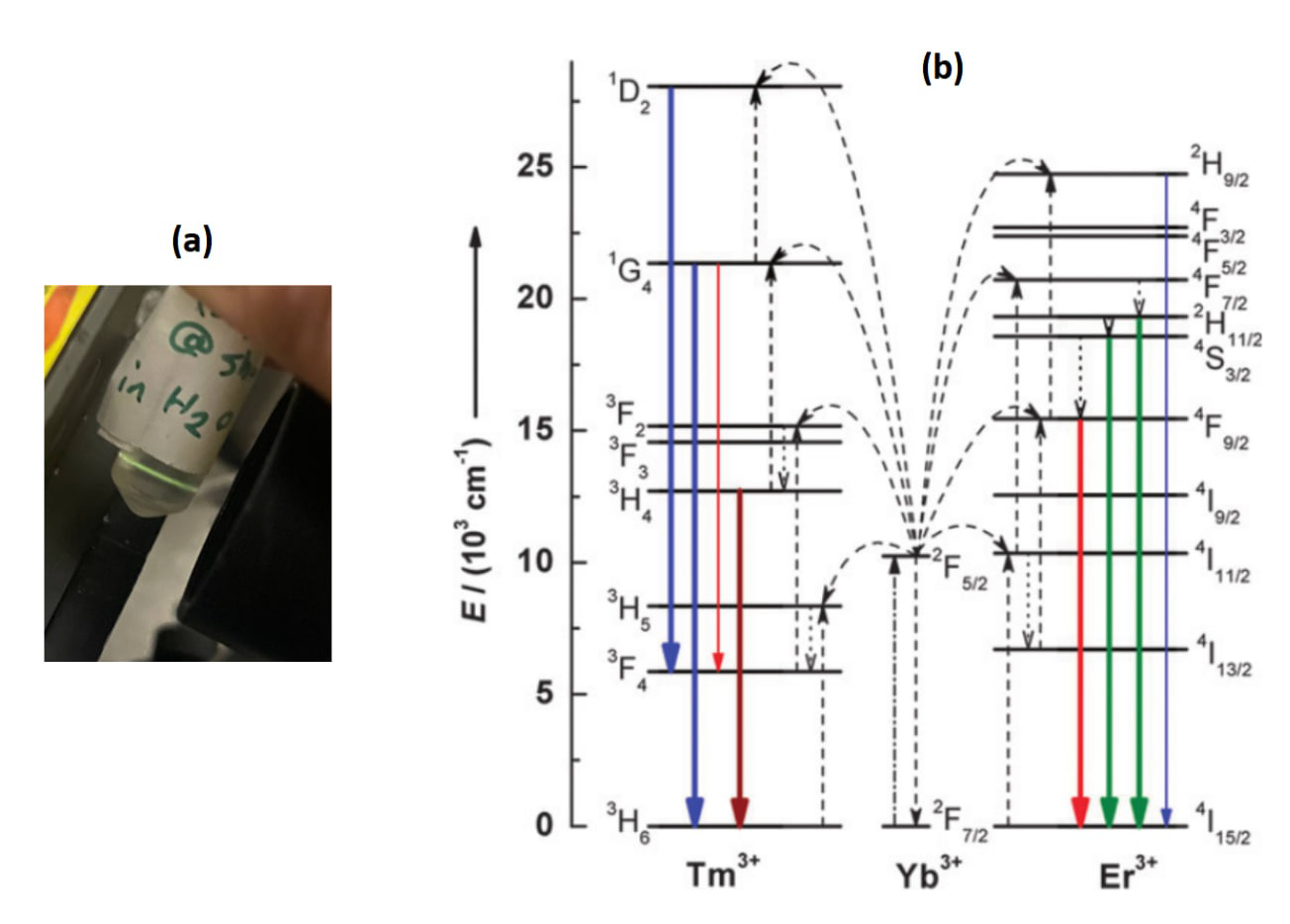}}
		\caption{(a): A vial contains a dispersion of the EBAA coated nanoparticles (LiYF$_4$:Yb(18\%), Er(1.5\%), Tm(0.5\%) dispersed in water under 980 nm excitation. (b): Energy transfer mechanisms involved in the upconversion processes in crystals doped with ${Er}^{3+}$, ${Tm}^{3+}$, and ${Yb}^{3+}$ under 980 nm laser excitation are illustrated. The dashed-dotted, dashed, dotted, and solid arrows represent photon excitation, energy transfer, phonon relaxation, and emission processes, respectively. The colors of the solid arrows correspond to the specific colors of the visible light emissions generated in each process. Adapted from \citep{wang2009recent}. }\label{fig10}
	\end{figure}

 The process involved heating a solution of 500 mg of EBAA  in  8 ml of diethylene glycol to 110 °C under argon flow with vigorous stirring. Then 1 ml (20 mg) of LiYF$_4$:Yb(18\%), Er(1.5\%), Tm(0.5\%) core particles dissolved in cyclohexene was injected into the flask. The hot solution turned cloudy right after a chloroform solution of UCNPs (20 mg) was injected. The mixture was maintained at a temperature of 240 °C for approximately 1.5 hours until a clear solution was obtained. The resulting solution is subjected to dialysis in diethylene glycol (DEG) for 20 hours to remove excess ligands and solvents. The hydrophilic UCNPs are then isolated via centrifugation, washed with pure water, and redispersed in deionized water for analysis. Figure (\ref{fig10}) shows the EBAA-coated nanoparticles (LiYF$_4$:Yb(18\%), Er(1.5\%), Tm(0.5\%)) dispersed in water and excited under a 980 nm diode laser.

\subsection{Characterization of Particle Size and Morphology: TEM and DLS Analysis}

\begin{figure}[H]
		\centerline{\includegraphics[width=1\textwidth,clip=]{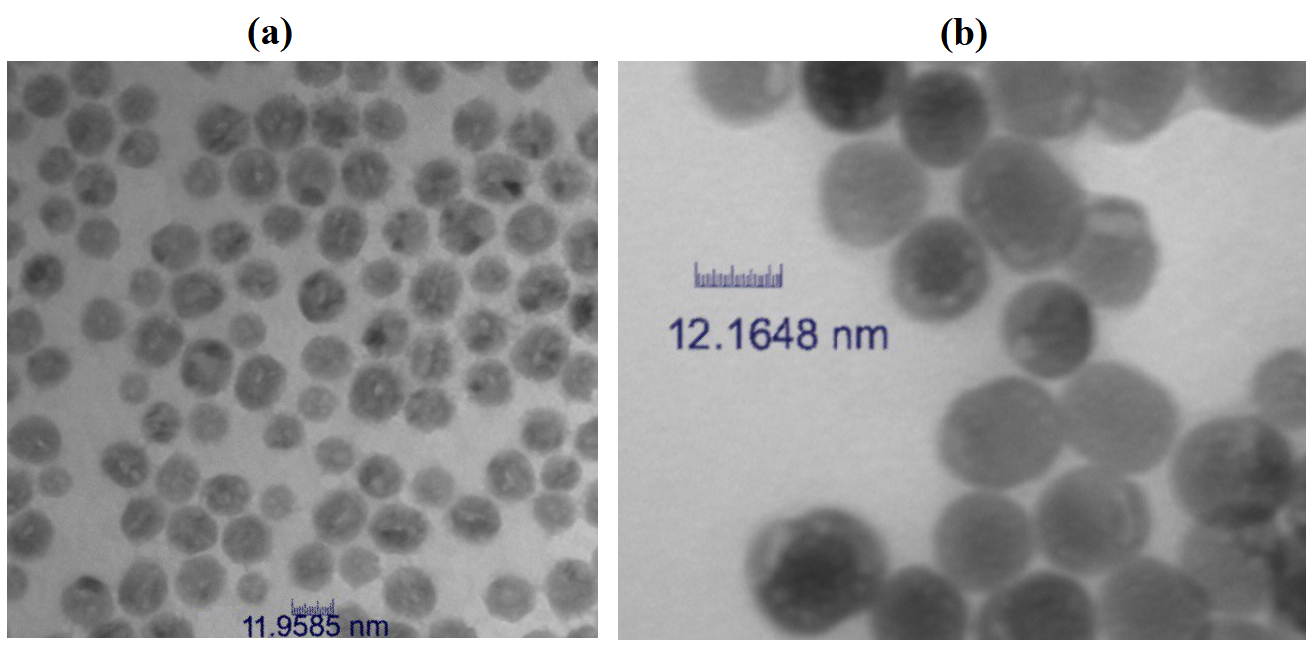}}
		\caption{TEM images of LiYF$_4$:Yb(18\%), Er(1.5\%), Tm(0.5\%) UCNPs before (a) and after (b) EBAA ligand exchange. Image (a) is captured at 100k magnification, and image (b) is captured at 200k magnification using TEM JEOL 1200.}\label{fig11}
	\end{figure}

The surface features of the samples were examined using transmission electron microscopy (TEM JEOL 1200). The size of both the original (oleate-coated) and EBAA-capped upconversion nanoparticles (UCNPs) was characterized dynamic light scattering (DLS). Figure (\ref{fig11}) presents TEM images of both original (oleate-coated) and EBAA-capped upconversion nanoparticles (UCNPs). As depicted in Fig. \ref{fig11}(a), the original UCNPs are spherical and exhibit near monodispersity, with an average diameter of approximately 12 nm. In comparison to the original UCNPs, the EDAA-capped UCNPs (shown in Fig. \ref{fig11}(b)) display no significant alterations in size or shape \citep{esmaeili2021monitoring}. 

\begin{figure}[H]
		\centerline{\includegraphics[width=1.17\textwidth,clip=]{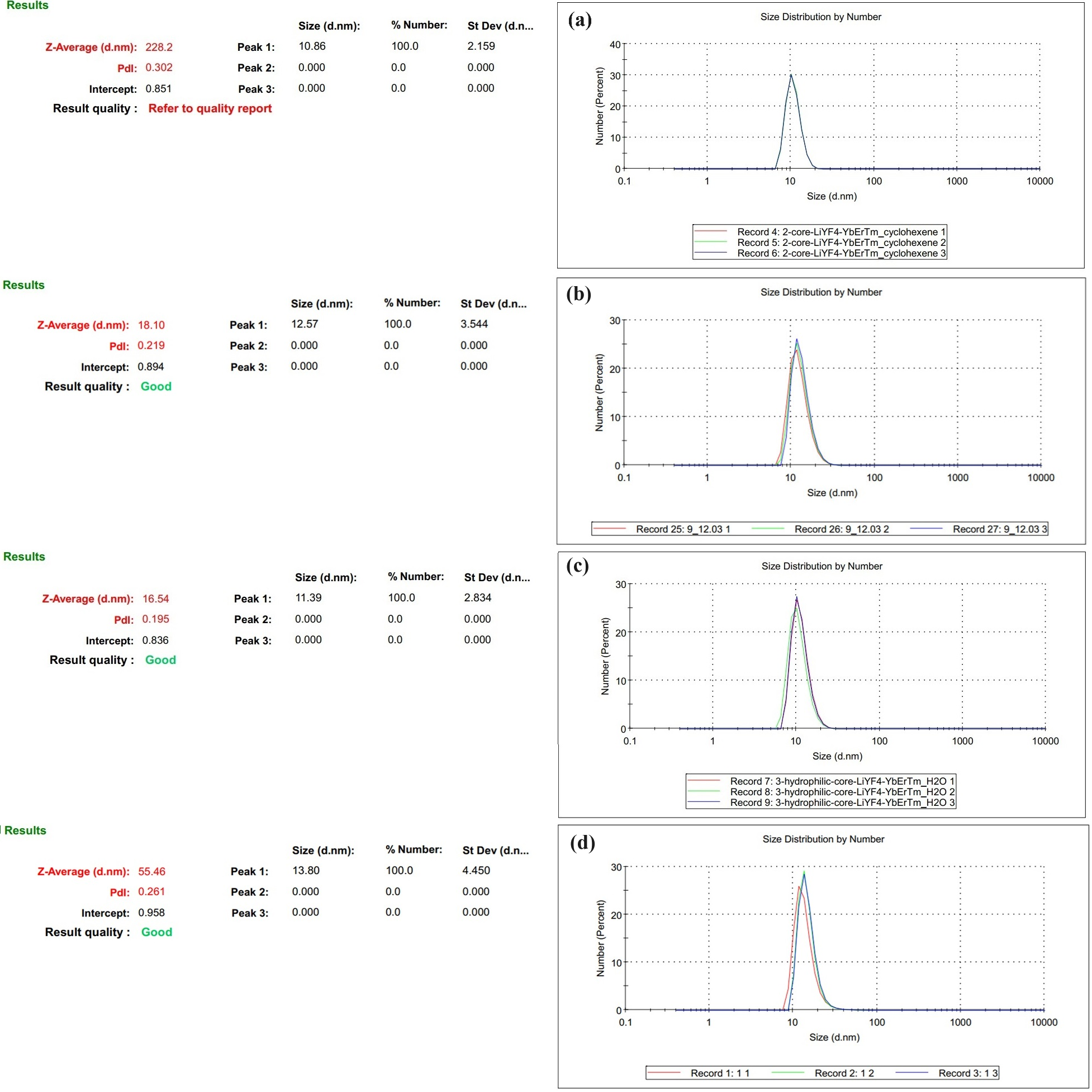}}
		\caption[DLS size distribution of various UCNPs samples]{(a) Original (oleate-coated) nanoparticles (LiYF$_4$:Yb(18\%), Er(1.5\%), Tm(0.5\%)) dispersed in cyclohexane, size: 10.86 nm. (b) EBAA-capped UCNPs (LiYF$_4$:Yb(18\%), Er(1.5\%), Tm(0.5\%)) dispersed in DIH$_2$O made on 05/11/2022, size: 12.57 nm. (c) EBAA-capped UCNPs (LiYF$_4$:Yb(18\%), Er(1.5\%), Tm(0.5\%)) dispersed in DIH$_2$O made on 05/22/2022, size: 11.39 nm. (d) EBAA-capped UCNPs (LiYF$_4$:Yb(18\%), Er(1.5\%), Tm(0.5\%)) dispersed in DIH$_2$O made on 05/22/2022 and 2nd DLS measured on 04/05/2023, size 13.8 nm.}\label{fig12}
	\end{figure}

Dynamic light scattering (DLS) was employed to complement the TEM analysis by providing insight into the size distribution of the nanoparticles. The DLS size distribution of the original (oleate-coated) nanoparticles (LiYF$_4$:Yb(18\%), Er(1.5\%), Tm(0.5\%)) dispersed in cyclohexane (Fig. \ref{fig12} (a)) and EBAA-capped UCNPs in water (Fig. \ref{fig12} (b)) demonstrated the uniformity and monodispersity of the particles. Additionally, Fig. \ref{fig12}(c) and \ref{fig12}(d) demonstrate that the stability of the size distribution remains consistent even after 348 days. Notably, the first DLS measurement on 05/22/2022 showed a size of 11.39 nm, and after 348 days on 04/05/2023, the same batch maintained a size of 13.8 nm, indicating excellent stability and no particle agglomeration over this extended period. These DLS results underscore the uniformity and monodispersity of the particles and suggest the success of surface modification, contributing to the colloidal stability and broad compatibility of the nanoparticles with various applications.

\subsection{Optical Characterization Setup}
The optical characteristics of the upconversion nanoparticles (UCNPs) were studied using a custom-made laser-scanning confocal microscope, as described in our previous works \citep{rajil2022quantum}. An oil immersion objective (Leica HCX Plan Apo 40×/1.25-0.75 OIL CS /0.17/E objective) was used to focus the 980nm laser (peak power 300mW) on the sample and collect the backscattered fluorescent light from the sample in an inverted configuration. The fluorescent light was passed through a transmission grating, and the spectrum was collected by an ICCD camera (Starlight Xpress Trius Pro 674). 

\begin{figure}[H]  
		\centerline{\includegraphics[width=1.1\textwidth,clip=]{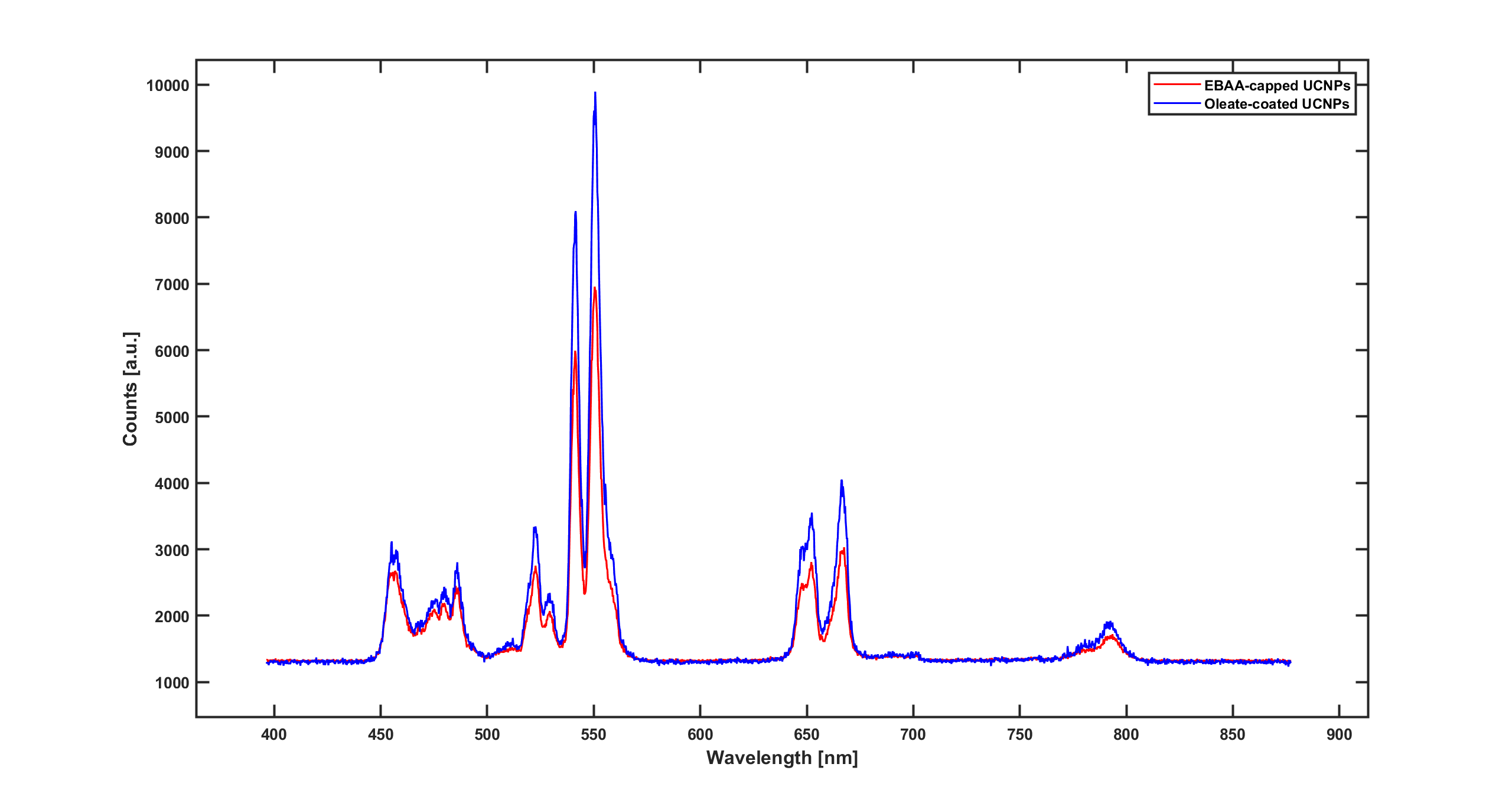}}
		\caption{Upconversion luminescence spectra (UCL) of the original oleate-coated LiYF$_4$:Yb(18\%), Er(1.5\%), Tm(0.5\%) nanoparticles dispersed in cyclohexane and EBAA-capped UCNPs in water, highlighting the emission properties of both types of nanoparticles under identical excitation conditions.}\label{fig13}
	\end{figure}

Following the development of our custom confocal laser-scanning microscope, we investigated the optical properties of both original oleate-coated LiYF$_4$:Yb(18\%), Er(1.5\%), Tm(0.5\%) nanoparticles dispersed in cyclohexane and EBAA-capped UCNPs in water. The upconversion luminescence spectra (UCL) presented in Fig. (\ref{fig13}) demonstrate the emission profiles of these nanoparticles under identical excitation conditions. The comparison of the emission spectra reveals the impact of the EBAA capping on the optical performance of the UCNPs,  notably by mitigating quenching effects. This analysis allows us to better understand the role of surface modification in improving the upconversion efficiency and potential applications of these nanoparticles in various biological and sensing systems.

\section{Comparative Analysis of the EBAA Method and Perspectives on Hydrophilic UCNPs}
\subsection{EBAA Method Comparison}

Several methods have been developed for the surface modification of UCNPs, each with its own advantages and limitations. In this comparison, we discuss the failures of some of the widely used surface modification methods and highlight the EBAA method as the best nominee to address these limitations. In this regard, we have meticulously reproduced each of the follwoing five surface modification methods for UCNPs three times to validate their performance. Through these repeated experiments, we were able to identify the specific limitations and failures associated with each method, which have informed our conclusion that the EBAA method is the most effective choice for addressing these challenges.

\begin{itemize}
\item \textbf{Lemieux-von Rudloff Reagent:}
The Lemieux-von Rudloff reagent is a method used for surface modification of UCNPs with carboxylic acid-functionalized groups, enabling their use as biological labels. In the study by \cite{chen2008versatile}, this method faced issues of precipitation and weak fluorescence, which limits its application in DNA-based biosensors. These limitations stem from the fact that the reagent can cause aggregation of the nanoparticles, resulting in reduced fluorescence intensity and poor colloidal stability in aqueous solutions.

\item \textbf{Carboxy-PEG4-Phosphonic Acid:}
The Carboxy-PEG4-phosphonic acid method is another surface modification technique used for upconverting NaYF$_4$ nanoparticles. This method involves the attachment of PEG-phosphate ligands to the nanoparticle surface, enhancing their compatibility with the biological window for near-infrared (NIR) biolabeling applications. However, as reported by \cite{boyer2010surface}, this method suffers from quenching issues, which can significantly reduce the fluorescence intensity of the UCNPs. Quenching is particularly problematic in DNA-based biosensors, where a strong signal is crucial for detecting the target analyte.

\item \textbf{Polyethylene Glycol 600 Diacid (HOOC–PEG–COOH):}
Polyethylene glycol 600 diacid (HOOC–PEG–COOH) is another method used for surface modification of hexagonal-phase NaYF$_4$:Yb, Er, and NaYF$_4$:Yb, Tm nanocrystals to improve their up-conversion fluorescence efficiency. As reported by \cite{yi2006synthesis}, this method has a major drawback in that particles tend to be lost during the washing steps at the end of the process. Losing a significant number of particles during the purification process can lead to reduced yields and negatively impact the overall performance of the modified UCNPs in DNA-based biosensors.

\item \textbf{Surface–Ligand Exchange of UCNPs by HDA:}
The surface–ligand exchange of UCNPs by hexanedioic acid (HDA) is a method used to transfer NaYF$_4$:Yb/Er (or Yb/Tm) up-converting nanoparticles from hydrophobic to hydrophilic environments. This process, as reported by \cite{zhang2009hexanedioic}, enables the UCNPs to be more compatible with aqueous environments and biological systems. However, this method suffers from quenching issues similar to those observed with the Carboxy-PEG4-phosphonic acid method, limiting its application in DNA-based biosensors.

\item \textbf{UCNP@SiO2:}
UCNP@SiO2 is a surface modification method that involves coating upconverting nanoparticles with a silica shell. This technique, as reported by \cite{melle2018forster} can improve biocompatibility, stability, and functionality for Förster resonance energy transfer (FRET) studies between UCNPs and quantum dots. However, this method has limitations due to the nonuniform silica shell, which may lead to inconsistencies in the performance of the modified UCNPs. The nonuniformity of the silica shell can introduce additional challenges in the dispersion and stability of the particles in biological systems. Furthermore, the uneven silica shell thickness may adversely affect the FRET efficiency between UCNPs and other fluorophores, such as quantum dots, by altering the distance between them. This could potentially result in inaccurate or unreliable measurements in optical sensing and biosensing applications.

Despite the advantages that the UCNP@SiO2 method brings, such as enhanced biocompatibility and stability, the nonuniform silica shell remains a critical limitation that needs to be addressed. In comparison, the EBAA method provides a uniform coating on UCNPs, eliminating issues related to FRET and ensuring optimal performance in various applications, including DNA-based biosensors, bioimaging, and sensing. Hence, while the UCNP@SiO2 method has its merits, the nonuniform silica shell presents a significant limitation that can affect the overall performance of the modified UCNPs in optical sensing and biosensing applications. The EBAA method, on the other hand, offers a more uniform and reliable surface modification approach, making it a superior choice for developing efficient and biocompatible upconverting nanoparticles for a wide range of applications.

\end{itemize}

Table (\ref{tab1}) provides a summary of upconversion nanoparticles (UCNPs) that have undergone surface modification through ligand engineering with carboxyl groups, along with their respective applications.

\begin{table}[h]
\caption{Comparison of Methods for Surface Modification of UCNPs}
\centering
\begin{tabularx}{\textwidth}{cXXX}
\toprule
\textbf{Method} & \textbf{Failure reason} & \textbf{Application} & \textbf{Reference} \\
\midrule
Lemieux-von Rudloff reagent & Precipitation and weak fluorescence & DNA-Based biosensor & \citep{chen2008versatile} \\
\addlinespace
Carboxy-PEG4-phosphonic acid & Quenching & deep tissue imaging & \citep{boyer2010surface} \\
\addlinespace
Polyethylene glycol 600 diacid (HOOC–PEG–COOH) & Lost particles during the washing steps & bioimaging, optical sensing & \cite{yi2006synthesis} \\
\addlinespace
Surface–ligand exchange of UCNPs by HDA & Quenching & biological labeling & \citep{zhang2009hexanedioic} \\
\addlinespace
UCNP@SiO2 & Nonuniform silica shell, silica shell may affect FRET & optical and biosensing & \citep{melle2018forster} \\
\addlinespace
\bottomrule
\end{tabularx}\label{tab1}
\end{table}

Hence, the carboxylate groups in EBAA replace the original hydrophobic ligands on the UCNPs' surface (oleate), rendering them hydrophilic. The EBAA method offers a promising approach for synthesizing hydrophilic UCNPs with enhanced properties that make them more suitable for a wide range of applications in aqueous environments. It is important to note that the EBAA-coated UCNPs exhibit remarkable long-term stability, with no signs of precipitation observed even after six months. Thus, the EBAA method offers improved stability, biocompatibility, and fluorescence properties for the synthesized hydrophilic UCNPs compared to existing surface modification techniques. The hydrophilic UCNPs synthesized using the EBAA method exhibit several advantageous properties compared to those produced by traditional surface modification techniques. These properties include:

\begin{itemize}
    \item \textbf{Size and shape control:} The EBAA method enables precise control over the size and morphology of the UCNPs, which can significantly impact their optical properties and biocompatibility.

\item \textbf{Surface charge:} The carboxylate groups introduced by EBAA provide the UCNPs with a negative surface charge, which can enhance their stability and reduce nonspecific adsorption in biological applications.

\item \textbf{Long-term stability:} The EBAA-coated UCNPs demonstrate exceptional long-term stability, with no precipitation observed even after six months. This attribute is highly desirable for applications that require extended shelf life and consistent performance over time.

\item \textbf{Optical properties:} Hydrophilic UCNPs synthesized using the EBAA method exhibit high fluorescence intensity and upconversion efficiency, owing to the reduced quenching effects associated with water and the absence of non-luminescent surface defects.

\item \textbf{Biocompatibility:} The use of EBAA as a surface modifier results in hydrophilic UCNPs with low cytotoxicity and excellent biocompatibility, making them suitable for various biomedical applications.

\end{itemize}

\subsection{Future Perspectives for Hydrophilic UCNPs}

The EBAA-synthesized hydrophilic UCNPs present a leap forward for applications ranging from bioimaging to energy conversion. Their enhanced fluorescence and stability improve imaging resolutions, while their biocompatibility opens doors to in vivo applications, including precise biosensing and targeted drug delivery. Innovations may soon yield UCNPs tailored for specific biological targets, multiplexed detection systems, and stimuli-responsive drug release mechanisms.

In energy conversion, these UCNPs could extend the capabilities of photovoltaic devices by harnessing NIR light, thus improving solar cell performance. Upcoming studies are likely to leverage the UCNPs’ distinctive features in developing new materials and device structures for optimized energy utilization. Furthermore, the broad potential of hydrophilic UCNPs is anticipated to expand into optogenetics and environmental monitoring, capitalizing on their unique fluorescence and compatibility to manipulate cellular functions and detect pollutants with unprecedented specificity and sensitivity

\section{Conclusion and Future Work}

In conclusion, our comparative study has meticulously evaluated the ethylene bis(2-aminoethyl)amine (EBAA) method against five other prevalent surface modification techniques for upconverting nanoparticles (UCNPs). This investigation has illuminated the challenges inherent in these methods, ranging from precipitation and weak fluorescence to quenching and nonuniform silica shell formation, which can negatively impact Förster resonance energy transfer (FRET). Contrasting these methods, the EBAA method emerges as a clear frontrunner, adeptly addressing the issues of weak fluorescence and quenching that plague other techniques. Its superior performance is evident in the fluorescence intensity, colloidal stability, yield, and uniform coatings it imparts to UCNPs. However, it is crucial to note that the outstanding performance of the EBAA method has been validated exclusively for LiYF$_4$: Yb, Er, Tm UCNPs. These UCNPs exhibit enhanced fluorescence intensity and colloidal stability, rendering them ideal for diverse applications, including bioimaging, drug delivery, and solar cells.

An empirical testament to the EBAA method's efficacy is the remarkable stability observed in LiYF$_4$: Yb, Er, Tm particles over an extended period. DLS measurements indicated a marginal size increase from 11.39 nm to 13.8 nm over 348 days, underscoring the absence of particle agglomeration and affirming their long-term application viability. As we look to the future, our research endeavors are pivoting towards exploring the adaptability of the EBAA method beyond LiYF$_4$: Yb, Er, Tm UCNPs. Preliminary efforts are being directed toward other classes of UCNPs, such as NaGdF$_4$ and NaYF$_4$. The objective is to determine if the EBAA method can replicate its efficacy with these particles, thereby validating its versatility and establishing a universal surface modification technique for diverse UCNP compositions.

This comprehensive exploration and comparison not only underscore the crucial need for selecting an appropriate surface modification technique but also stimulates continued research and enhancement of UCNPs for a variety of applications. By diligently refining and adapting the EBAA method, we aim to drive advancements in upconverting nanoparticle technology, setting the stage for innovative breakthroughs in bioimaging, optical sensing, and biosensing.

\section*{Acknowledgements}
This research was supported by multiple esteemed organizations, and we extend our sincere gratitude to each for their invaluable contributions. We acknowledge the financial support from the Air Force Office of Scientific Research (Award No. FA9550-20-1-0366), Office of Naval Research (Award No. N00014-20-1-2184), and the Robert A. Welch Foundation (Grants No. A-1261, A-1547). Our work also benefited from grants from the National Science Foundation (Grant No. PHY-2013771, PHY-1820930, ECCS-2032589), and the National Institutes of Health (Award No. R03AI139650 and R21AI149383).

This material is based upon work supported by the U.S. Department of Energy, Office of Science, Office of Biological and Environmental Research under Award Number DE-SC-0023103. Special acknowledgment is given to the Hagler Institute for Advanced Study at Texas A\&M University for their support of SE's involvement in this research. Additionally, we thank the Herman F. Heep and Minnie Belle Heep Texas A\&M University Endowed Fund, held and administered by the Texas A\&M Foundation, which supported AH's contributions. Each organization's dedication to promoting scientific research has been crucial to our endeavors, and we are profoundly thankful for their support.
	

	\bibliographystyle{abbrvnat}
	\setcitestyle{authoryear}
	\bibliography{library}
\end{document}